\documentclass[12pt]{article}
\usepackage{geometry,epstopdf}
\usepackage{graphicx}

\def \d {{\rm d}} 
\def \boldu {\mbox{\boldmath$u$}}

\begin{document}

\title{\bf A rotating cylinder in an asymptotically locally anti-de~Sitter background} 

\author{J. B. Griffiths$^1$\thanks{E--mail: {\tt J.B.Griffiths@lboro.ac.uk}} \ 
and N. O. Santos$^{2,3}$\thanks{E--mail: {\tt N.O.Santos@qmul.ac.uk}}
\\ \\ \small
$^1$Department of Mathematical Sciences, Loughborough University, \\
\small Loughborough,  Leics. LE11 3TU, U.K.\\ 
\small $^2$Department of Mathematical Sciences, Queen Mary University of London,\\
\small Mile End Road, London, E1 4NS, U.K. \\
\small $^3$Laborat\'orio Nacional de Computa\c{c}\~ao Cient\'{\i}fica, 25651-070 Petr\'opolis RJ, Brazil. } 

\date{\today}
\maketitle

\begin{abstract}
\noindent 
A family of exact solutions is presented which represents a rigidly rotating cylinder of dust in a background with a negative cosmological constant. The interior of the infinite cylinder is described by the G\"odel solution. An exact solution for the exterior solution is found which depends both on the rotation of the interior and on its radius. For values of these parameters less than a certain limit, the exterior solution is shown to be locally isomorphic to the Linet--Tian solution. For values larger than another limit, it is shown that the exterior solution extends into a region which contains closed timelike curves. At large distances from the source, the space-time is shown to be asymptotic locally to anti-de~Sitter space. 
\end{abstract}

%\newpage 
%\bigskip\bigskip
\section{Introduction}

Bonnor, Santos and MacCallum~\cite{BoSaMC98} have identified a family of compound space-times for which the interior of an infinite cylinder of finite radius is described by the G\"odel solution~\cite{Godel49}. They showed that this can be matched to a vacuum exterior in a form obtained by Santos~\cite{Santos93}. The resulting space-time is cylindrically symmetric and stationary. It claims to describe the interior and exterior fields of a rigidly rotating cylinder of dust in a vacuum background with a negative cosmological constant. However, the exterior space-time was not expressed explicitly and some properties were only outlined. It is the purpose of this paper to construct this solution explicitly in a form that is suitable for interpretation. In this way, the properties of the space-time are identified for all values of the characterising parameters.

After identifying an appropriate form of the G\"odel metric that is to be used as an internal solution, we present a modified form of metric for the family of solutions that are considered for the external region. We then apply the appropriate junction conditions and identify expressions for the parameters for which it is matched to a G\"odel interior. Some properties of the space-time are then described in detail.

%\newpage
\section{The G\"odel solution} 

The perfect-fluid solution obtained by G\"odel~\cite{Godel49} can be expressed in comoving coordinates in the cylindrical form 
 \begin{equation} 
 \d s^2= -\d t^{\,2} +\d\rho^2+\d z^2 
 -{2\sqrt2\over\omega}\sinh^2\!\omega\rho\,\d\phi\,\d t
 +{1\over\omega^2}\!\left( \sinh^2\!\omega\rho-\sinh^4\!\omega\rho\right)\!\d\phi^2, 
 \label{Godelmetric2} 
 \end{equation} 
 where $\omega$ is an arbitrary constant, ${\rho\in[0,\infty)}$, ${\phi\in[0,2\pi)}$ and ${\phi=2\pi}$ is identified with ${\phi=0}$. The coordinate $\rho$ may be interpreted as the proper distance from a regular axis at ${\rho=0}$, and the rotational symmetry of the metric~(\ref{Godelmetric2}) about the axis can be clearly seen. The fluid four-velocity is ${\boldu=\partial_t}$, and the constant density~$\mu$ (the pressure~${p=0}$) and cosmological constant~$\Lambda$ are given by 
 $$ 8\pi\mu=4\omega^2, \qquad \Lambda=-2\omega^2. $$ 
 This implies that the fluid density is positive and the cosmological constant is negative. The Weyl tensor is of type~D and there are no curvature singularities. This form of the metric has been preferred here since it explicitly reduces to a vacuum Minkowski space in the slowly rotating limit as ${\omega\to0}$.

This solution, which is clearly stationary, is characterised by the fact that it is spatially homogeneous and the fluid has no expansion, acceleration or shear. The only non-zero kinematic quantity is the {\em rotation} (vorticity), which has constant magnitude~$\omega$ that is aligned with the axis. Its properties are well known (see e.g.~\cite{Godel49} and \cite{HawEll73}--\cite{GriPod09}).

This solution may be interpreted as describing a homogeneous space-time with a fluid source that is rigidly rotating. In the Newtonian analogue of such a situation, relative to any point, there will exist a cylinder of finite radius around the axis of rotation on which the particles of the fluid would move at the speed of light. For all points outside this cylinder, the particles of the classical fluid will be moving faster than light. Such a situation is not possible in a relativistic theory but this feature is reflected in the appearance of {\em closed timelike curves}.

In fact, it can be seen from the metric (\ref{Godelmetric2}) that curves on which $t$, $\rho$ and $z$ are constant are either spacelike or timelike according to whether ${\sinh^2\omega\rho}$ is, respectively, less than or greater than~1. They are null when ${\omega\rho=\log(1+\sqrt2)}$ (or ${\sinh(2\omega\rho)=2\sqrt2}$). Moreover, since ${\phi=2\pi}$ is identified with ${\phi=0}$, these are closed timelike curves whenever $\rho$ is larger than this value, but they are not geodesics. In the present paper, we are considering an infinite cylinder described by this metric and so it is assumed that, in the interior region, ${\omega\rho<\log(1+\sqrt2)}$.

\section{Possible exterior solutions} 

We now consider a family of compound space-times which contain an infinite rigidly-rotating cylinder of matter with proper radius~$\rho_1$, such that ${\omega\rho_1<\log(1+\sqrt2)}$. The region inside this cylinder is taken to be described by the G\"odel metric~(\ref{Godelmetric2}) with ${0<\rho<\rho_1}$. The exterior region is taken to be vacuum with, of course, a negative cosmological constant. Our purpose is both to find the exterior solution which matches to~(\ref{Godelmetric2}) appropriately on the boundary ${\rho=\rho_1}$, and to determine its properties.

The exterior metric may be assumed to take the general stationary form 
 \begin{equation} 
 \d s^2 =  -A\,\d t^2 -2K\,\d t\,\d\phi +\d\rho^2 +C\,\d\phi^2 +D\,\d z^2 ,
 \label{GenMetric}
 \end{equation} 
 where $\rho$ remains as a proper radial distance and $A$, $C$, $D$ and $K$ are functions of~$\rho$ only. It is assumed here that ${t\in(-\infty,\infty)}$, ${\rho\in(0,\infty)}$, ${z\in(-\infty,\infty)}$ and ${\phi\in[0,2\pi)}$. With ${\phi=2\pi}$ identified with ${\phi=0}$, it may be seen that the space-time contains closed timelike curves whenever the metric function $C$ is negative. In this case, $\phi$~is a timelike coordinate and curves with constant $t$, $\rho$ and~$z$ are closed timelike curves, though they are not necessarily geodesics.

It is also convenient to introduce a function ${\Delta(\rho)}$ such that 
 \begin{equation} 
 \Delta^2=AC+K^2, 
 \label{Deltasquare} 
 \end{equation} 
 and it is reasonable to assume that ${\Delta>0}$. With this, the metric (\ref{GenMetric}) could be written in the alternative form 
  $$ \d s^2 =  -A\left(\d t+{K\over A}\,\d\phi\right)^2 +\d\rho^2 
 +{\Delta^2\over A}\,\d\phi^2 +D\,\d z^2. $$

\subsection{The BSM exterior solution} 

In~\cite{BoSaMC98}, Bonnor, Santos and Mac\-Callum have considered the exterior solution to be a special case of the family of vacuum solutions given by Santos~\cite{Santos93}. Using a slightly different notation involving the above constant ${\omega=\sqrt{-\Lambda/2}}$, this can be expressed in terms of two functions ${G(\rho)}$ and ${F(\rho)}$, where ${G=\sqrt{-g}=\Delta\sqrt{D}}$.

In particular, they have taken $G$ in the form 
 \begin{equation} 
 G=c_1\cosh\Big(\sqrt6\,\omega\rho\Big)
 +c_2\sinh\Big(\sqrt6\,\omega\rho\Big), 
 \label{Gdef1} 
 \end{equation} 
 where $c_1$ and $c_2$ are constants, and have required $F(\rho)$ to satisfy the equation 
 \begin{equation} 
 F'=\gamma\,G^{-1}, 
 \label{FGeqn} 
 \end{equation} 
 where $\gamma$ is a constant. With these, the metric function $D$ takes the form 
 $$ D=\epsilon\,\Delta\,e^{\delta F}, $$ 
 where $\epsilon$ and $\delta$ are real constants with ${\epsilon>0}$. The function $\Delta$ is then given by 
 \begin{equation} 
 \Delta=\epsilon^{-1/3}\,G^{2/3}\,e^{-\delta F/3}, 
 \label{Deltadef}
 \end{equation} 
 and the remaining metric functions were taken to be 
 \begin{eqnarray} 
 &&A=\Delta\left[2\beta\cosh\alpha F 
 -\Big({\alpha^2+\beta^2\over\alpha}\Big)\sinh\alpha F\right], \nonumber\\ 
 &&K=\Delta\left[\cosh\alpha F 
 -{\beta\over\alpha}\sinh\alpha F\right], \label{Santossol}\\ 
 &&C=\Delta\,{1\over\alpha}\sinh\alpha F, \nonumber 
 \end{eqnarray} 
 which contain the additional constants $\alpha$ and $\beta$, of which $\alpha$ could be imaginary. The parameters of this solution satisfy the constraint 
 \begin{equation} 
 24\,\omega^2({c_2}^2-{c_1}^2) =(3\alpha^2+\delta^2)\gamma^2. 
 \label{constraint} 
 \end{equation}

On application of the boundary conditions, however, it is found that this form of the exterior solution does not easily reduce to Minkowski space in the slowly rotating limit as ${\omega\to0}$. A different approach is therefore preferred here.

\subsection{A general exterior solution}

In order to construct an exterior solution that more transparently contains the slowly rotating limit, we take the same solution as given by (\ref{Gdef1})--(\ref{Deltadef}) and (\ref{constraint}) but apply a general transformation to the $t$ and $\phi$ coordinates. In this way, in place of (\ref{Santossol}), the remaining metric functions can be expressed in the more general form 
 \begin{eqnarray} 
 && A= \Delta \left[ 2\kappa\mu\cosh\alpha F
 -(\kappa^2+\mu^2)\sinh\alpha F\right], \nonumber\\[5pt] 
 && K= \Delta \left[ (\kappa\nu-\lambda\mu)\cosh\alpha F
 +(\kappa\lambda-\mu\nu)\sinh\alpha F\right], \\[5pt] 
 && C= \Delta \left[ 2\lambda\nu\cosh\alpha F
 +(\lambda^2+\nu^2)\sinh\alpha F\right], \nonumber 
 \end{eqnarray} 
 where $\kappa$, $\lambda$, $\mu$ and $\nu$ are arbitrary constants which satisfy the constraint 
 \begin{equation} 
 \kappa\nu+\lambda\mu=1, 
 \label{kappalambdaconstraint} 
 \end{equation} 
 and are such that $\kappa$ and $\mu$ have dimension $L^{-1/2}$ and $\lambda$ and $\nu$ have dimension $L^{1/2}$. Even with the condition (\ref{kappalambdaconstraint}), however, this parametrisation contains more freedom than is required. A further constraint may be imposed. However, this will be left until after the application of junction conditions and it will then be used in order to simplify the expressions obtained.

\goodbreak
Our second point of departure from~\cite{BoSaMC98}, is to note that it is possible to determine the exterior solution explicitly by first rewriting the constants $£c_1$ and~$c_2$ as 
 $$ c_1=-{c\over\sqrt6\,\omega}\sinh(\sqrt6\,\omega\rho_0), \qquad
 c_2={c\over\sqrt6\,\omega}\cosh(\sqrt6\,\omega\rho_0), $$ 
 where $c$ and $\rho_0$ are constants. This re-expresses the function (\ref{Gdef1}) in the form 
 \begin{equation} 
 G={c\over\sqrt6\,\omega}\sinh\Big(\sqrt6\,\omega(\rho-\rho_0)\Big). 
 \label{Gcase1} 
 \end{equation} 
 In this case, the equation (\ref{FGeqn}) can be integrated to give $F(\rho)$ explicitly in the form 
 \begin{equation} 
 F= {\gamma\over c}\, \log\left[ {2k\over\sqrt6\,\omega} 
 \tanh\left({\sqrt6\,\omega\over2}(\rho-\rho_0) \right) \right], 
 \label{Fcase1} 
 \end{equation} 
 where $k$ is an arbitrary constant. Recalling that ${\omega=\sqrt{-\Lambda/2}}$, it is now convenient to introduce the functions 
 \begin{equation} 
 P(\rho)={2\tanh\left(\sqrt{-3\Lambda}(\rho-\rho_0)/2\right)
 \over\sqrt{-3\Lambda}}, \qquad 
 Q(\rho)={\sinh\left(\sqrt{-3\Lambda}(\rho-\rho_0)\right)
 \over\sqrt{-3\Lambda}}, 
 \label{PQ} 
 \end{equation} 
 so that ${G=c\,Q}$ and ${e^{F}=(kP)^{\gamma/c}}$. This metric could have a singularity when ${\rho=\rho_0}$. However, when taken to represent the exterior of a cylinder of proper radius~$\rho_1$, this does not occur for ${\rho\in[\rho_1,\infty)}$.

In terms of the above functions,  
 \begin{equation} 
 \Delta= \epsilon^{-1/3}\,(c\,Q)^{2/3}\,(kP)^{-\gamma\delta/3c}, 
 \label{Delta} 
 \end{equation} 
 and the remaining metric functions are given explicitly by 
 \begin{eqnarray} 
 &&A= {\textstyle{1\over2}} \,\Delta\left[ -(\kappa-\mu)^2\,(kP)^{\alpha\gamma/c} 
 +(\kappa+\mu)^2\,(kP)^{-\alpha\gamma/c} \right], \label{Adef}\\[5pt] 
 &&K= {\textstyle{1\over2}} \,\Delta\left[ (\kappa-\mu)(\lambda+\nu)\,(kP)^{\alpha\gamma/c} 
 -(\kappa+\mu)(\lambda-\nu)\,(kP)^{-\alpha\gamma/c} \right], \\[5pt] 
 &&C= {\textstyle{1\over2}} \,\Delta\left[ (\lambda+\nu)^2\,(kP)^{\alpha\gamma/c} 
 -(\lambda-\nu)^2\,(kP)^{-\alpha\gamma/c} \right], \\[6pt] 
 &&D= \Delta\,\epsilon\,(kP)^{\gamma\delta/c}, \label{Ddef} 
 \end{eqnarray} 
 in which the constants satisfy the constraint 
 \begin{equation} 
 (3\alpha^2+\delta^2)\gamma^2=4c^2, 
 \label{altcondition} 
 \end{equation} 
 replacing~(\ref{constraint}). The above expressions remain valid when $\alpha$ and $c$ are either real or imaginary. However, when $c$ is imaginary, $\rho_0$ is complex.

%\newpage
\section{Matching conditions} 

It is now appropriate to consider the compound space-time in which an infinite cylinder of finite proper radius~$\rho_1$, whose interior is described by the G\"odel metric, is matched to an exterior described by the metric in the family described above.

The interior G\"odel solution in the form (\ref{Godelmetric2}) is characterised by a single parameter $\omega$, which determines both the density of the dust source and the value of the cosmological constant. A cylinder of such a source is also determined by its proper radius~$\rho_1$. A rigidly rotating cylinder described in this way is thus characterised by these two parameters.

Such a source should now be matched to a vacuum exterior solution described by the metric~(\ref{GenMetric}) in its general form with (\ref{Adef})--(\ref{Ddef}) and (\ref{PQ}) and~(\ref{Delta}). It may first be noted that the metrics (\ref{Godelmetric2}) and (\ref{GenMetric}) for each region are expressed in terms of the same proper radial coordinate~$\rho$. Then, according to the Lichnerowicz conditions, the metric functions and their derivatives must be continuous across the junction at ${\rho=\rho_1}$. This provides eight explicit constraints on the parameters in the exterior metric. Given that the parameters $\omega$ (and hence~$\Lambda$) and~$\rho_1$ are determined by the source, the eight junction conditions determine the values of the quantities $c$, $\rho_0$, $\epsilon$, $\gamma\delta$, $\alpha\gamma$, $\kappa$, $\lambda$, $\mu$, $\nu$ and $k$, in which $\kappa$, $\lambda$, $\mu$ and $\nu$ are subject to (\ref{kappalambdaconstraint}) and some additional constraint that is yet to be identified.

It can immediately be seen that the continuity of ${G=\sqrt{-g}}$ and its derivative on the junction ${\rho=\rho_1}$ implies the two conditions 
 \begin{eqnarray} 
 \sqrt2\,c\,\sinh\Big(\sqrt6\,\omega(\rho_1-\rho_0)\Big) 
 &=& \sqrt3 \sinh(2\omega\rho_1), \nonumber\\ 
 c\,\cosh\Big(\sqrt6\,\omega(\rho_1-\rho_0)\Big) &=& \cosh(2\omega\rho_1). \nonumber 
 \end{eqnarray} 
 Thus $c$ is given by 
 \begin{equation} 
 c^2=1-2\sinh^2\omega\rho_1-2\sinh^4\omega\rho_1, 
 \label{ccase1} 
 \end{equation} 
 and $\rho_0$ is defined by the equation 
 \begin{equation} 
 \sqrt2\,\tanh\left(\sqrt6\,\omega(\rho_1-\rho_0)\right) 
 =\sqrt3\,\tanh\left(2\omega\rho_1\right). 
 \label{rho0def3}
 \end{equation} 
 Thus ${Q_1\equiv Q(\rho_1)}$, and hence ${P_1\equiv P(\rho_1)}$, are now determined explicitly as 
 $$ c\,Q_1={\sinh(2\omega\rho_1)\over2\omega}, \qquad
 P_1={2\Big(\cosh(2\omega\rho_1)-c\Big)\over3\omega\>\sinh(2\omega\rho_1)}. $$

It may also be noted that equations (\ref{rho0def3}) and (\ref{ccase1}), which define $\rho_0$ and $c$, only have real solutions when 
 $$ 0<\omega\rho_1<\sinh^{-1}\sqrt{\sqrt3-1\over2}\quad (<0.573108), $$ 
 or when ${0<\sinh(2\omega\rho_1)<\sqrt2}$. For values of $\omega\rho_1$ outside this range, $c$ is imaginary and $\rho_0$ complex.

\goodbreak 
Now consider the (assumed positive) quantity $\Delta$ defined by (\ref{Deltasquare}). The continuity of this and its derivative across ${\rho=\rho_1}$ leads to the two further constraints 
 \begin{equation} 
 \epsilon={2\omega\over\sinh(2\omega\rho_1)}\,(kP_1)^{-\delta\gamma/c}, \qquad
 \gamma\delta =-\cosh(2\omega\rho_1). 
 \label{epsdelgam} 
 \end{equation} 
 These two conditions are equivalent to the required continuity of $D$ and its derivative.

\goodbreak
Now consider the continuity of $A/\Delta$, $K/\Delta$, $C/\Delta$ and their derivatives. These six (not independent) conditions can be expressed as 
 \begin{eqnarray} 
 &&\hskip-3pc-(\kappa-\mu)^2\,(kP_1)^{\alpha\gamma/c} 
 +(\kappa+\mu)^2\,(kP_1)^{-\alpha\gamma/c} 
 = {2\omega\over\sinh\omega\rho_1\cosh\omega\rho_1}, \label{Acond}\\[4pt] 
 &&\hskip-3pc(\kappa-\mu)(\lambda+\nu)\,(kP_1)^{\alpha\gamma/c} 
 -(\kappa+\mu)(\lambda-\nu)\,(kP_1)^{-\alpha\gamma/c} 
 = 2\sqrt2\>{\sinh\omega\rho_1\over\cosh\omega\rho_1}, \label{Kcond}\\[4pt] 
 &&\hskip-3pc(\lambda+\nu)^2\,(kP_1)^{\alpha\gamma/c} 
 -(\lambda-\nu)^2\,(kP_1)^{-\alpha\gamma/c} 
 = {2\sinh\omega\rho_1(1-\sinh^2\omega\rho_1)\over\omega\>\cosh\omega\rho_1}, \label{Ccond}\\[4pt] 
 &&\hskip-3pc-(\kappa-\mu)^2\,(kP_1)^{\alpha\gamma/c} 
 -(\kappa+\mu)^2\,(kP_1)^{-\alpha\gamma/c} 
 = -{2\omega\over\alpha\gamma}\>{(1+2\sinh^2\omega\rho_1)\over\sinh\omega\rho_1\cosh\omega\rho_1}, \label{Aprimecond} \\[4pt] 
 &&\hskip-3pc (\kappa-\mu)(\lambda+\nu)\,(kP_1)^{\alpha\gamma/c} 
 +(\kappa+\mu)(\lambda-\nu)\,(kP_1)^{-\alpha\gamma/c}
 = {2\sqrt2\over\alpha\gamma}\>{\sinh\omega\rho_1\over\cosh\omega\rho_1}, \label{Kprimecond} \\[4pt] 
 &&\hskip-3pc(\lambda+\nu)^2\,(kP_1)^{\alpha\gamma/c} 
 +(\lambda-\nu)^2\,(kP_1)^{-\alpha\gamma/c}    
 = {2\sinh\omega\rho_1(1-3\sinh^2\omega\rho_1-2\sinh^4\omega\rho_1)\over
\alpha\gamma\omega\>\cosh\omega\rho_1}. \label{Cprimecond} 
 \end{eqnarray}

The combinations 
(\ref{Ccond})(\ref{Kprimecond})$-$(\ref{Kcond})(\ref{Cprimecond}), 
(\ref{Ccond})(\ref{Aprimecond})$-$(\ref{Acond})(\ref{Cprimecond}) and 
(\ref{Kcond})(\ref{Aprimecond})$-$(\ref{Acond})(\ref{Kprimecond}) with the constraint (\ref{kappalambdaconstraint}) give, respectively, 
 \begin{eqnarray} 
 \alpha\gamma(\lambda^2-\nu^2) 
 &=& {2\sqrt2\over\omega}\>\sinh^4\omega\rho_1, \nonumber \\ 
 \alpha\gamma(\kappa\lambda+\mu\nu) &=& 1-2\sinh^2\omega\rho_1, \label{genconds} \\[4pt] 
 \alpha\gamma(\kappa^2-\mu^2) &=& 2\sqrt2\>\omega. \nonumber 
 \end{eqnarray} 
 These equations can be solved by putting 
 $$ \kappa=A\cosh\psi, \qquad \lambda=B\cosh\chi, \qquad \mu=A\sinh\psi, \qquad \nu=B\sinh\chi, $$ 
 where, according to (\ref{kappalambdaconstraint}), the constants $A$, $B$, $\psi$ and $\chi$ are subject to the constraint 
 $$ AB\sinh(\psi+\chi)=1. $$ 
 The conditions (\ref{genconds}) then become 
 $$ A^2={2\sqrt2\>\omega\over\alpha\gamma}, \qquad
 B^2={2\sqrt2\over\alpha\gamma\omega}\>\sinh^4\omega\rho_1, \qquad 
 AB\cosh(\psi+\chi)={1-2\sinh^2\omega\rho_1 \over\alpha\gamma}. $$ 
 Together, these imply that 
 \begin{equation} 
 \alpha^2\gamma^2 =1-4\sinh^2\omega\rho_1-4\sinh^4\omega\rho_1. 
 \label{alphagamma} 
 \end{equation} 
 It follows that $\alpha$ is only real for 
 $$ 0<\omega\rho_1<\sinh^{-1}\sqrt{\sqrt2-1\over2}\quad (<0.440687), $$ 
 i.e.~for ${0<\sinh(2\omega\rho_1)<1}$. 
 Significantly, it can now be seen that the expressions (\ref{ccase1}), (\ref{epsdelgam}) and (\ref{alphagamma}) satisfy the required condition (\ref{altcondition}).

\goodbreak
At this point, it can be seen to be convenient to use the remaining freedom in the choice of the parameters $\lambda$, $\kappa$, $\mu$ and~$\nu$, to put ${\chi=\psi}$ (so that ${\kappa\nu=\lambda\mu={1\over2}}$). With this additional assumption it can then be seen that 
 $$ \sinh^2\psi={1-2(1+\sqrt2)\sinh^2\omega\rho_1\over4\sqrt2\sinh^2\omega\rho_1}, \qquad
 \cosh^2\psi={1+2(\sqrt2-1)\sinh^2\omega\rho_1\over4\sqrt2\sinh^2\omega\rho_1}, $$ 
 and thus 
 \begin{eqnarray} 
 (\kappa+\mu)^2={\omega(1-2\sinh^2\omega\rho_1+\alpha\gamma)
 \over\alpha\gamma\>\sinh^2\omega\rho_1}, \quad&& 
 (\lambda+\nu)^2={\sinh^2\omega\rho_1(1-2\sinh^2\omega\rho_1+\alpha\gamma)
 \over\alpha\gamma\omega}, \nonumber\\ 
 (\kappa-\mu)^2={\omega(1-2\sinh^2\omega\rho_1-\alpha\gamma)
 \over\alpha\gamma\>\sinh^2\omega\rho_1}, \quad&& 
 (\lambda-\nu)^2={\sinh^2\omega\rho_1(1-2\sinh^2\omega\rho_1-\alpha\gamma)
 \over\alpha\gamma\omega}, \nonumber
 \end{eqnarray}

%\goodbreak 
So far, seven independent junction conditions and two constraints have been used to obtain explicit expressions for the quantities $c$, $\rho_0$, $\epsilon\,k^{\gamma\delta/c}$, $\gamma\delta$, $\alpha\gamma $, $\kappa$, $\lambda$, $\mu$ and $\nu$. One more condition is required to separately identify the parameters $\epsilon$ and~$k$. Any of the conditions \hbox{(\ref{Acond})--(\ref{Cprimecond})} or combinations of them is sufficient for this. In particular, subtracting (\ref{Aprimecond}) from (\ref{Acond}) gives 
 $$ (kP_1)^{\alpha\gamma/c} = 
 {\cosh\omega\rho_1\>(1+\alpha\gamma-2\sinh^2\omega\rho_1) \over
 \sinh\omega\rho_1\>(1+\alpha\gamma+2\sinh^2\omega\rho_1)}. $$ 
 This determines~$k$, and $\epsilon$ is then obtained from (\ref{epsdelgam}).

\section{The slowly rotating limit} 

It is now appropriate to investigate the limit in which ${\omega\to0}$ while $\rho_1$ is kept constant. In this limit, in which the cosmological constant vanishes, the metric (\ref{Godelmetric2}) for the interior region approaches that of a vacuum Minkowski space in cylindrical coordinates. It therefore needs to be checked that the exterior metric approaches the same limit.

Expanding expressions to the least order in $\omega$ that is required, it can be seen that ${c\to1}$, ${\rho_0\to0}$, ${\gamma\delta\to-1}$, ${\epsilon\to(\rho_1kP_1)^{-1}}$, ${c\,Q_1\to\rho_1}$, and hence 
 $$ \Delta\to \rho_1\bigg({Q\over Q_1}\bigg)^{2/3}\bigg({P\over P_1}\bigg)^{1/3} \to\rho.  $$ 
 It can then be seen that 
 $$ \alpha\gamma\to1-2\omega\rho_1-{\textstyle{14\over3}}(\omega\rho_1)^2 +\dots $$ 
 Hence 
 $$ (\kappa+\mu)^2\to{2\omega\over(\omega\rho_1)^2}, \quad
 (\kappa-\mu)^2\to4\omega(\omega\rho_1)^2, \quad 
 (\lambda+\nu)^2\to{2\over\omega}(\omega\rho_1)^2, \quad
 (\lambda-\nu)^2\to{4\over\omega}(\omega\rho_1)^6. $$ 
 Since ${kP_1\to(\omega\rho_1)^{-1}}$ and ${P\to\rho}$, it follows that ${k\to\omega^{-1}\rho_1^{-2}}$. It can then be seen that the remaining metric functions approach the following limits: 
 $$ A\to1, \qquad K\to0, \qquad C\to\rho^2. $$ 
 It is thus confirmed that the exterior metric reduces exactly to the required form of Minkowski space in this limit.

\section{Asymptotic properties} 

It is also appropriate to investigate the asymptotic properties of this family of solutions as ${\rho\to\infty}$. It may first be noted from (\ref{Gcase1}) that, in this limit, 
 $$ G=c\,Q\to{c\over2\sqrt6\,\omega}e^{\sqrt6\,\omega(\rho-\rho_0)}. $$ 
 Moreover, $P$ approaches the finite quantity ${\sqrt2/(\sqrt3\,\omega)}$. It can then be seen from (\ref{Delta}) that $\Delta$ approaches a constant multiplied by ${e^{2\sqrt6\,\omega\rho/3}}$. Thus, $A$, $K$, $C$ and $D$ all behave asymptotically like ${e^{2\sqrt6\,\omega\rho/3}}$ and it is possible to rotate coordinates to set ${K\to0}$. Then, after appropriately rescaling the coordinates, the metric asymptotically approaches a form that is equivalent to 
 \begin{equation} 
 \d s^2 =\exp\bigg(2\sqrt{-{\Lambda\over3}}\,\rho\bigg)(-\d\bar t^2 
 +\d\bar z^2 +\d\bar\phi^2) +\d\rho^2, 
 \label{adSmetric1} 
 \end{equation} 
 which is one of the standard forms for the metric of anti-de~Sitter space. This  confirms that this solution is also asymptotic to the known static cylindrically symmetric solution with a negative cosmological constant \cite{Linet86}, \cite{Tian86}. However, it should be noted that the coordinate $\bar\phi$ in (\ref{adSmetric1}) is periodic. The exterior region is therefore not strictly asymptotic to anti-de~Sitter space in any global sense.

\section{The exterior region} 

It can now be shown that the coordinate transformation 
 \begin{equation} 
 t={1\over\sqrt2}\Big[(\lambda+\nu)\tilde t-(\lambda-\nu)\tilde\varphi\Big], \qquad \phi={1\over\sqrt2}\Big[(\kappa-\mu)\tilde t-(\kappa+\mu)\tilde\varphi\Big], 
 \label{transtostat} 
 \end{equation} 
 takes the vacuum exterior metric (\ref{GenMetric}), with components given by (\ref{Adef})--(\ref{Ddef}), to the diagonal form 
 $$ \d s^2=\Delta\Big[ -(kP)^{-\alpha\gamma/c}\d\tilde t^{\,2}
 +\epsilon(kP)^{\gamma\delta/c}\d z^2 
 +(kP)^{\alpha\gamma/c}\d\tilde\varphi^2\Big] +\d\rho^2, $$ 
 where $\Delta$ is given by (\ref{Delta}). It is now possible to rescale the coordinates $\tilde t$, $\tilde\varphi$ and $z$ to absorb the constants $\epsilon$, $c$ and $k$, so that the resulting metric becomes 
 \begin{equation} 
 \d s^2= Q^{2/3}\Big[ -P^{-(3\alpha+\delta)\gamma/3c}\d\bar t^2
 +P^{2\delta\gamma/3c}\d\bar z^2 
 +P^{(3\alpha-\delta)\gamma/3c}\d\bar\varphi^2 \Big] +\d\rho^2. 
 \label{locstatmetric} 
 \end{equation} 
 This may now be seen to be identical to the general static metric with a cosmological constant obtained by Linet \cite{Linet86} and Tian \cite{Tian86}, which may be expressed in the form 
 \begin{equation} 
 \begin{array}{l} 
 \d s^2=Q^{2/3}\Big[ -P^{-2(1-8\sigma+4\sigma^2)/3\Sigma}\,\d t^2 
 +P^{-2(1+4\sigma-8\sigma^2)/3\Sigma}\,\d z^2 \\ 
 \hskip14pc +P^{4(1-2\sigma-2\sigma^2)/3\Sigma}\,\d\varphi^2 \Big] 
 +\d\rho^2, 
 \end{array} 
 \label{LinetTianmetric} 
 \end{equation} 
 where ${\Sigma=1-2\sigma+4\sigma^2}$, and in which $Q$ and $P$ are given exactly as in (\ref{PQ}) for this case with ${\Lambda<0}$. (The periods of $\bar\varphi$ and $\varphi$ and the existence of related conicity parameters are ignored in both metrics.) 
 Notice that the sum of the different powers of~$P$ in~(\ref{LinetTianmetric}) vanishes, while the sum of the squares of the powers of~$P$ is equal to~${8\over3}$. For the metric~(\ref{locstatmetric}), the first result is satisfied trivially and the second is precisely the condition~(\ref{altcondition}).

The Linet--Tian metric (\ref{LinetTianmetric}) is the static generalisation of the Levi-Civita solution which includes a cosmological constant. The parameter~$\sigma$ is usually interpreted as the mass per unit length of the source as it has this interpretation in the limit as ${\Lambda\to0}$. By comparing the indices in the metrics (\ref{locstatmetric}) and (\ref{LinetTianmetric}), it can be seen that the parameter~$\sigma$ is given by 
 $$ \sigma ={1\over4} \mp{\sqrt3\over4}\sqrt{1+{\gamma\delta\over2c}\over1-{\gamma\delta\over2c}}. $$ 
 Thus ${\sigma=0}$ corresponds to ${\gamma\delta=-c}$ which, as expected, occurs when ${\omega\rho_1=0}$ and the source vanishes.

The results given here indicate that the exterior field in {\em locally} static. However, it is not globally static. In addition, it only has this property when the expressions in the transformation (\ref{transtostat}) are real, and this only occurs when $\alpha$ is real. The situation here is therfore qualitatively the same as that described by the van~Stockum solution \cite{vanStockum37}, which describes a rigidly rotating cylinder of dust when ${\Lambda=0}$ and in which, provided ${\omega\rho_1}$ is not too large, the vacuum exterior region is locally isomorphic to the Levi-Civita solution.

%\newpage 
\section{Conclusion} 

The solution described above represents the gravitational field a rotating infinite cylinder of finite radius. The interior region, given by the G\"odel metric~(\ref{Godelmetric2}), contains a rigidly rotating dust source. This has been matched to a vacuum exterior. Both regions have the same negative cosmological constant. The complete space-time is determined by two parameters -- the vorticity~$\omega$ of the fluid interior, which determines both the cosmological constant and the density of the fluid in the interior region, and $\rho_1$, which is the proper radius of the cylinder.

When ${0<\sinh(2\omega\rho_1)<1}$ (i.e.~when ${0\le\omega\rho_1<0.441}$ approximately), $\alpha$ is real and the exterior solution is locally isomorphic to the static Linet--Tian solution. When ${1<\sinh(2\omega\rho_1)<\sqrt2}$, $\alpha$ is imaginary and $c$ and $\rho_0$ are real. For ${\sqrt2<\sinh(2\omega\rho_1)}$ (i.e.~when ${0.573<\omega\rho_1}$ roughly), $\alpha$ and $c$ are both imaginary and $\rho_0$ is complex.

\begin{figure}[ht]
\begin{center} 
\includegraphics[scale=0.9, trim=5 5 5 5]{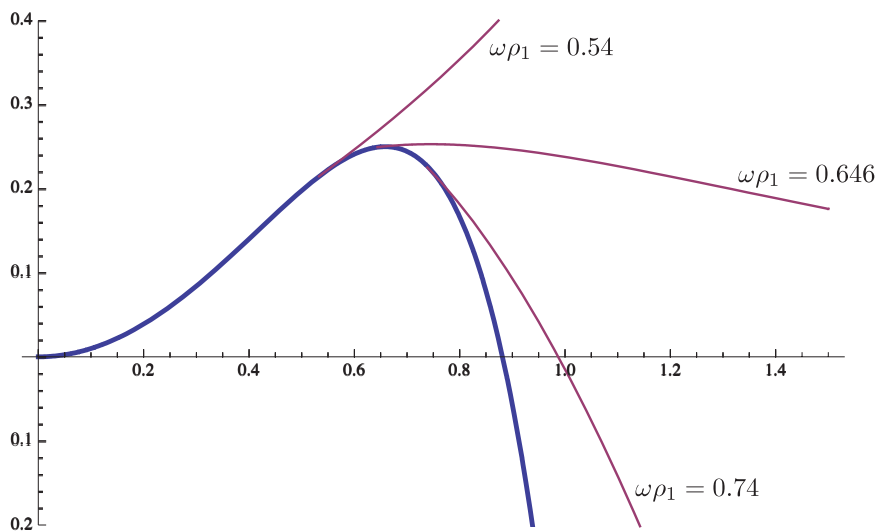} 
\end{center}
\caption{\small The heavy line illustrates the metric function $g_{\phi\phi}$ of the G\"odel solution as a function of $\omega\rho$. This is considered for ${\omega\rho\le\omega\rho_1}$. Extensions to the exterior solution are also illustrated for three particular values of~$\omega\rho_1$. Closed timelike curves occur when $g_{\phi\phi}$ is negative. For ${0<\omega\rho_1<0.646}$, the extension remains positive for all~$\rho$. For ${0.646<\omega\rho_1<0.881}$, the extension becomes negative after some finite distance and, for ${0.881<\omega\rho_1}$, the metric function is already negative in the interior region.  }
\label{Cfunction}
\end{figure}

The space-time contains closed timelike lines whenever the metric function~$C$ becomes negative. This function is plotted in figure~\ref{Cfunction} for extensions at three different values of~$\omega\rho_1$. It is found numerically that, if ${0<\omega\rho_1<0.6456}$, the exterior region does not contain closed timelike curves. For ${0.6456<\omega\rho_1<0.881}$, the exterior region extends after a finite distance into a region which contains closed timelike curves. For ${2\sqrt2<\sinh(2\omega\rho_1)}$, closed timelike curves occur in both the interior and exterior regions.

\section*{Acknowledgements} 

The authors are most grateful to Dr F\'atima da~Silva for hospitality at the 
Departamentode F\'{\i}sica Te\'orica, Universidade do Estado do Rio de Janeiro, Brazil, where this work was started. They are also grateful for financial support from CNPq, CAPES, Coordena\c{c}\~ao de P\'os Gradua\c{c}\~ao em F\'{\i}sica da UERJ and the Santander Fund.

\end{document}